# Robust class-gated single-pixel diffractive optical neural network with random-aberration-aware training

Xianjin Liu[1,2], Qiwen Bao[1,2], Ting Ma[1,2], Yihuan Liang[1,2], Yongqiu Lai[1,2], Bolun Zhang[1,2], Fansanqiu Li[1,2], Licheng Wang[1,2], and Jun-Jun Xiao[1,2,*]

[1]College of Integrated Circuits, Harbin Institute of Technology (Shenzhen), Shenzhen 518055, China.

[2]Shenzhen Engineering Laboratory of Aerospace Detection and Imaging, Harbin Institute of Technology (Shenzhen), Shenzhen 518055, China.

[*]E-mail: eiexiao@hit.edu.cn

**Optical computing offers the theoretical potential for high-speed, energy-efficient inference, yet its practical deployment remains constrained by fundamental input-output bottlenecks—particularly the reliance on electronic sensors with limited frame rates and stringent alignment requirements between optical components. Here, we demonstrate an image-class–gated single-pixel DONN that overcomes these limitations by converting spatial complexity into a temporal intensity signature. Using a minimal architecture comprising a reconfigurable digital micromirror device and a single-pixel photodetector, we implement a virtual optical gate. The system time-multiplexes class-specific masks, causing the detector response to peak only when the mask index matches the input class. This allows the predicted label to be read out via peak timing rather than spatial localization, eliminating 2D sensor constraints. To bridge the persistent sim-to-real gap, we introduce a physics-aware training strategy using random-phase augmentation. This method renders the model intrinsically tolerant to phase aberrations and mechanical misalignments without requiring precise hardware modeling. Our prototype achieves 90.0% (MNIST) and 80.0% (Fashion-MNIST) accuracy at a readout rate of 5 kHz. By combining gigahertz-compatible single-pixel detection with robust and alignment-tolerant training, this work provides a scalable, hardware-efficient pathway toward real-time optical intelligent sensing.**

## INTRODUCTION

In recent years, artificial intelligence has catalyzed transformative advances across computer vision, autonomous driving, and scientific discovery (1–5). The rapid evolution of large-scale models, exemplified by systems such as ChatGPT (6), is reshaping productivity and knowledge workflows. However, this progress comes at a steep cost: the growing complexity of models has triggered an exponential increase in computational and energy demands. In the post-Moore's-law era (7), conventional electronic architectures are fundamentally limited by the von Neumann bottleneck and I/O bandwidth constraints, making it difficult to achieve both high throughput and high energy efficiency. This has spurred intensive research into next-generation computing paradigms. Among them, optical computing stands out by harnessing the inherent parallelism of light propagation, low power dissipation, and ultra-high speed, offering a compelling hardware pathway toward high-throughput, low-latency intelligent systems (8–14).

Diffractive optical neural networks (DONNs), which perform computations through free-space wavefront modulation using cascaded diffractive layers, have emerged as a promising optical computing platform (15–24). Since their first demonstration in 2018 (16), DONNs have shown strong potentials in tasks ranging from image classification (16, 25–28) and logic operations (29–32) to quantitative phase imaging (33–35), optical encryption (36, 37) and optical mode multiplexing for high-capacity communication (38, 39), and generative applications (40, 41). However, their deployment faces three critical challenges. First, competitive performance typically requires multiple spatially cascaded layers, which are highly sensitive to inter-layer alignment. Even pixel level misregistration can cause severe accuracy loss (42). Second, a persistent simulation-to-real gap exists because parameters optimized in digital models often degrade under physical imperfections (43–45) such as device-specific phase aberrations. While hardware-in-the-loop or in-situ optimization can reduce this gap, they introduce significant optomechanical complexity and time overhead (45–47). Third, the operational bandwidth of DONNs is ultimately limited by the readout sensor. Array detectors (e.g., CMOS/CCD cameras) are framerate limited and expensive in non-visible bands, whereas point-scanning methods severely constrain throughput. Consequently, although light propagation occurs at the speed of light, the end-to-end system throughput is often bottlenecked by slow spatial readout.

To address these issues, we introduce and experimentally validate a class-gated single-pixel diffractive neural network (Gated-SP-DONN). Our design converts spatial decision-making into a temporal intensity signature, thereby eliminating the need for multi-region spatial readout and relaxing alignment constraints. The system employs minimalist architecture consisting of a single reconfigurable digital micromirror device (DMD) and a fixed single-pixel photo detector. By time-multiplexing trained, class-specific diffractive masks, the detector response peaks only when the mask index matches the input class. The predicted label is thus read out via peak timing rather than spatial localization. To bridge the sim-to-real gap without precise hardware modeling, we further introduce a random-phase-augmentation (RPA) training strategy that injects random phase perturbations during training, rendering the model intrinsically robust to phase aberrations and detection-area misalignments.

We demonstrate the prototype on MNIST and Fashion-MNIST classification, achieving experimental accuracies over 90.0% and 80.0%, respectively, using a single-pixel photodiode (PD) readout, in close agreement with numerical simulations. By combining gigahertz-compatible single-pixel detection with hardware-tolerant training, this work provides a scalable, hardware-efficient pathway toward real-time optical intelligent sensing, paving the way for the practical deployment of high-speed diffractive neural networks.

## RESULTS

### Design and Operation Principle of the Gated Single-Pixel DONN

Figure 1 compares the proposed Gated-SP-DONN with a conventional multi-layer DONN that uses spatial multi-region readout. The traditional DONN implements stepwise spatial modulation by cascading multiple diffractive layers and perform intelligent inference by reading and comparing optical intensities across multiple regions on the output plane (see Fig. 1a). In contrast, the proposed minimalist architecture (Fig. 1b) consists of only a single reconfigurable diffractive layer and a fixed output plane. Using a high-speed reconfigurable modulator, we sequentially display $c$ trained class-specific masks, each held for a fixed dwell time $\Delta T$. A single-pixel photodetector (PD) records the resulting temporal waveform at a sampling rate matched to the system bandwidth. The masks are optimized such that for an input image belonging to the $l$-th class, the optical intensity in the detection region peaks during display of the $c$-th mask, while remaining low during other mask intervals and during background periods. Classification is thus achieved by identifying the time window containing the maximum intensity. Conceptually, this gating mechanism functions as a set of class-indexed virtual optical gates that concentrate energy into the detector only when the gate index matches the input class. This approach reduces hardware complexity and, when combined with high-bandwidth modulators and PDs, can substantially increase computational throughput.

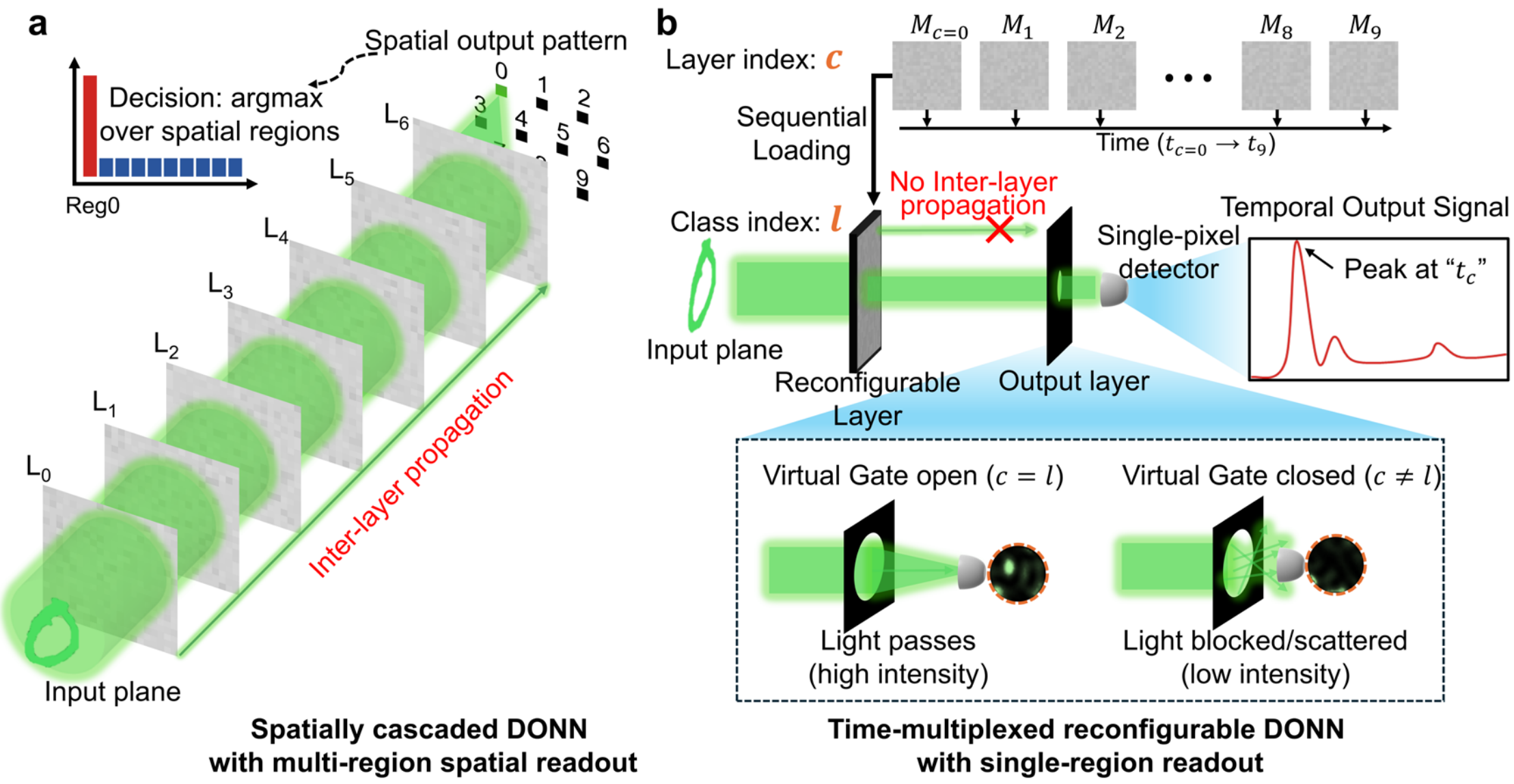

**Figure 1. Operational principle of the image-class-gated single-pixel diffractive neural network (Gated-SP-DONN). a.** Architecture of conventional multi-layer DONN: multiple diffractive layers are spatially cascaded to progressively modulate the incident optical field; inference is performed by integrating and comparing optical intensities across predefined detection regions on the output plane. **b.** Gated single-pixel DONN: a minimalist configuration with a single reprogrammable diffractive modulation plane and a fixed single-pixel detector. Note that $c$ trained class-specific masks are displayed sequentially with uniform dwell time $\Delta T$. For an input of class $l$, the detector intensity is significantly enhanced during the $c$-th time window and remains relatively much lower for non-matching windows and background; classification is obtained by identifying the time window containing the maximum intensity.

As a proof-of-concept, we employ a DMD operated in binary-amplitude mode as the reconfigurable diffractive layer, taking advantage of its high switching speed which matches the sampling bandwidth of the single-pixel detector. Figure 2a illustrates the system architecture and workflow. For each input image $I(x_i, y_i)$, the system combines it sequentially with all class-specific masks $M_c(x_i, y_i)(c = 0, \dots, 9)$ via an element-wise (Hadamard) multiplication, generating ten composite patterns $C_c(x_i, y_i)$. These patterns are displayed successively on the DMD (see Supplementary Note 1), each for a uniform dwell time $\Delta T$, while a single-pixel detector positioned at a fixed distance records the temporal signal. The class label is determined by locating the time bin containing the maximum intensity peak. To account for intrinsic phase aberrations of the DMD, we measured its phase profile using a four-step phase-shifting method; the reconstructed profile is shown in Fig. 2b (see Supplementary Note 2). Incorporating this measured phase into the forward model substantially reduced the discrepancy between simulation and experiment. Figure 2c shows the experimental setup, with key parameters and settings provided in **Methods**.

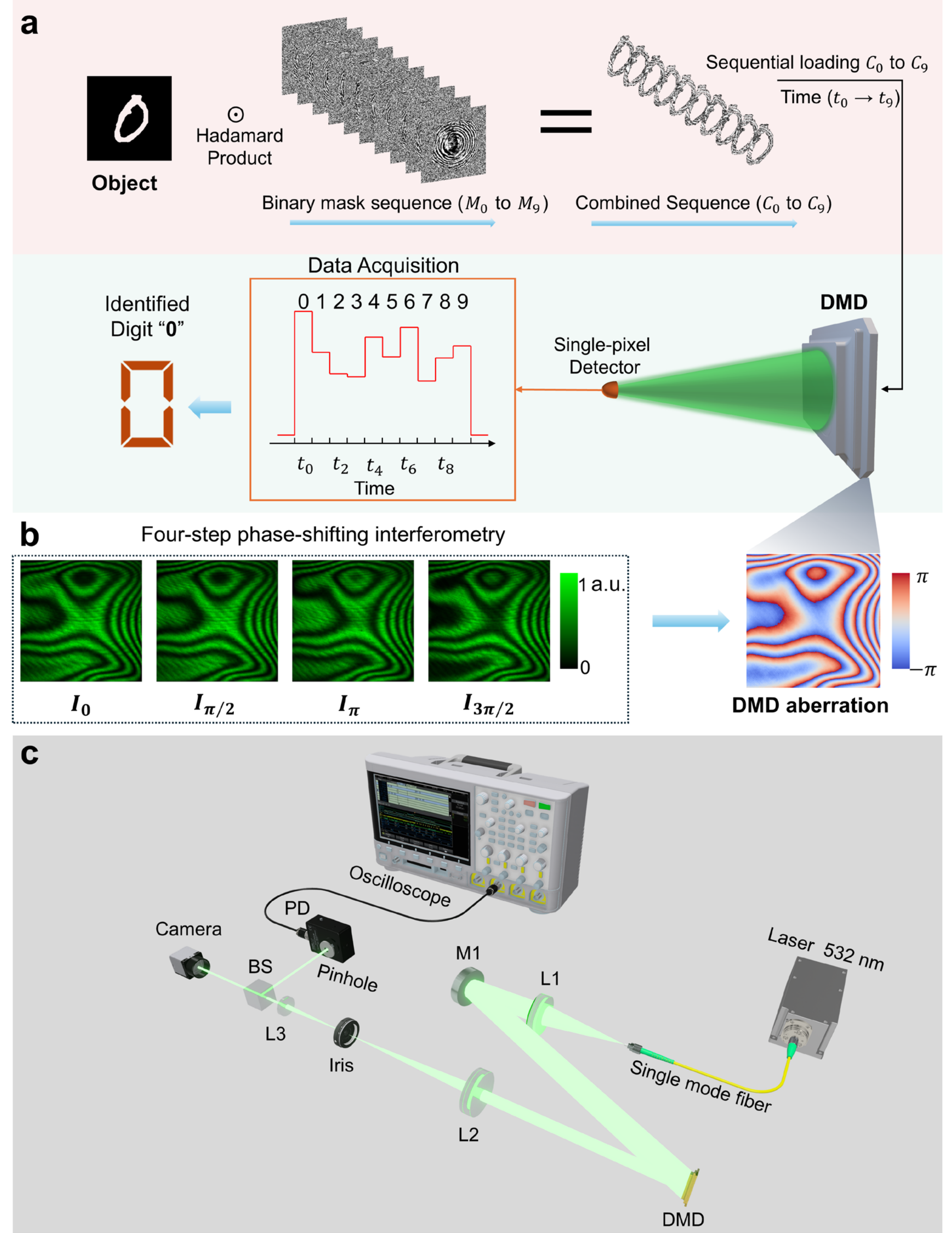

**Figure 2. Operational framework and experimental implementation of the gated diffractive neural network. a**. **System workflow**. For each input image $I(x_i, y_i)$, the system sequentially combines it with the ten class-specific masks $M_c(x_i, y_i)$ via element-wise (Hadamard) multiplication, generating ten composite patterns $C_c(x_i, y_i)$. These patterns are displayed sequentially on the DMD, each held for a uniform dwell time $\Delta T$. A single-pixel detector records the resulting temporal waveform, and the predicted class is obtained from the time bin exhibiting the maximum intensity. Key parameters are provided in Methods. **b. Phase calibration.** Representative interferograms acquired using a four-step phase-shifting method and the reconstructed DMD phase profile. **c. Experimental setup.** L1, plano-convex lens; M1, mirror; L2 and L3, achromatic lenses; BS, beamsplitter; PD, silicon photodiode.

## Experimental Validation and Sensitivity Analysis

We constructed an experimental prototype for handwritten digit classification on the MNIST dataset. Two models were compared: (1) an idealized model that considers device is perfect, and (2) a hardware-informed model that incorporated the measured DMD phase profile. Initially, a monochrome CMOS camera was used as a readout device to characterize the output field and locate the detection regions precisely. For each input, the camera captured ten output fields corresponding to the ten composite patterns; the predicted class was determined by integrating the intensity within each predefined detection region and selecting the frame with the highest response (see Methods).

Figure 3a shows representative classification results for the input image of digit "0". The ten composite patterns displayed on the DMD are shown in the top row. The corresponding simulated and experimental output plane intensity distributions are displayed in the middle and bottom rows, respectively. Additional representative simulation and experimental results for the detection regions of all ten MNIST digits are provided in Supplementary Note 3. The close agreement between simulation and experiment confirms the high modeling fidelity of the minimalist architecture. Figure 3b shows the integrated intensity distribution across the ten detection regions for this sample, verifying correct identification of the digit "0". We systematically evaluated the impact of hardware non-idealities on the final performance, focusing on two primary error sources: (1) intrinsic device phase aberration and (2) positional misregistration of the detection region. Experimental results indicate that the DMD's phase aberrations significantly limit experimental accuracy. When parameters trained with an idealized model were deployed on hardware, accuracy dropped from 95% in simulation to 85% experimentally. Incorporating the measured DMD phase profile into the forward model preserved 95% simulation accuracy and increased experimental accuracy to 92%, underscoring the necessity of explicitly modeling hardware-induced phase errors.

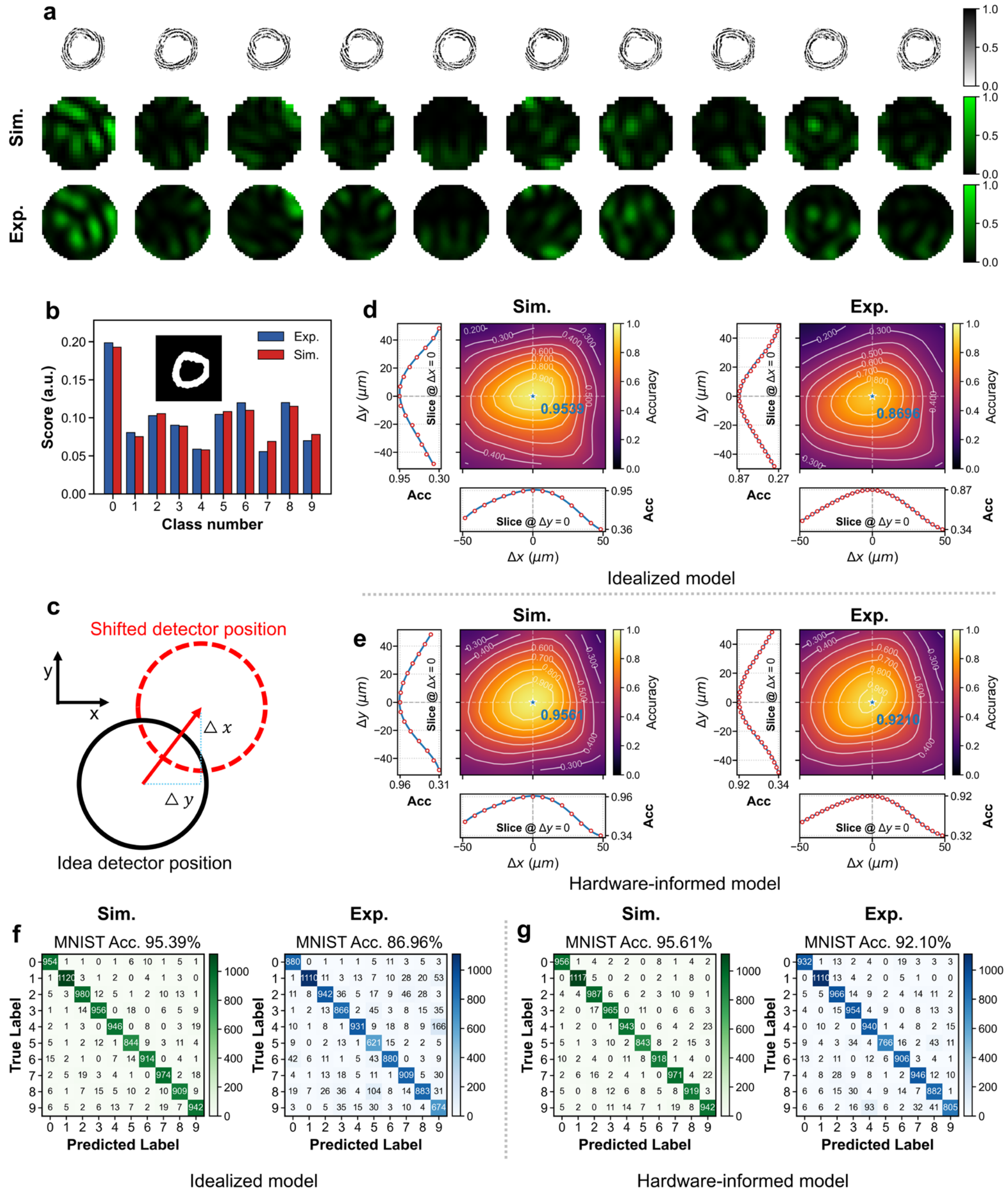

**Figure 3. Experimental validation and sensitivity analysis of the single-pixel gated diffractive network**. **a.** Representative classification results for the digit "0": the ten composite patterns displayed sequentially on the DMD (top) and the corresponding output-plane intensity distributions from simulation (middle) and experiment (bottom). **b.** Integrated intensity responses across the ten detection regions for the same sample; the inset shows the input digit. **c.** Definition of detection-region misregistration. $\Delta x$ and $\Delta y$ denote the relative lateral offset of the detection-region center with respect to the nominal optical axis. **d,e.** Classification accuracy versus detection-region offset $(\Delta x, \Delta y)$ for the ideal model (**d**) and the hardware-informed model (**e**) incorporating the measured phase-aberration profile. **f,g.** Confusion matrices for the idealized and hardware-informed models, respectively, evaluated at their optimal reference coordinates. The hardware-informed model achieves 92% experimental accuracy, compared with 86% for the idealized model, while both maintain accuracy 95% in simulation.

Beyond phase distortions, classification accuracy proved highly sensitive to detection region misalignment. The detection region was aligned using a reference hologram (Supplementary Note 4), but pixel-grid discretization introduced residual pixel-level misregistration. To compensate for this deviation, we performed a pixel-level translational scan of the output plane and selected the coordinates yielding maximal accuracy as the reference position. In the offset-sensitivity evaluation, spatial displacements $\Delta x$ and $\Delta y$ relative to this reference were used to quantify positional error (see Fig. 3c). Accuracy dropped sharply with increasing $|\Delta x|$ and $|\Delta y|$, revealing a narrow tolerance window. Figures 3d and 3e show the dependence of classification accuracy on detection-area offset for the idealized model (without counting the hardware-associated error sources discussed above) and hardware-informed model (counting the hardware associated error sources), respectively. Both exhibit a consistent degradation: when the offset approaches ~36.5% of the detection-area diameter, accuracy falls to ~32%, rendering the system nearly inoperative. Cut lines along $\Delta x = 0$ and $\Delta y = 0$ further illustrate the narrow tolerance. Experimentally, a deviation of only 10% of the detection-area diameter along either axis reduces the accuracy below 80%. This positional sensitivity is even more pronounced on the more complex Fashion-MNIST dataset (Supplementary Note 5). Figures 3f and 3g present the confusion matrices for the two models at their respective optimal coordinates. In summary, device-induced phase aberration and detection-area positional offset are the primary factors degrading performance in the single-pixel gated readout.

**Robustness by Training via Random Phase Perturbations**

As discussed above, although the gated diffractive network achieves effective classification, its sensitivity to phase aberrations and detection-area misalignment limits real-world deployment due to lack of robustness. To address this, we propose a RPA strategy. Instead of relying on the fixed hardware-informed model, RPA injects random phase perturbations during training, enabling the network to develop generalized compensation capabilities. As shown in Fig. 4a, random phase perturbations with spatial-frequency statistics matching the fluctuation scale of the measured DMD aberrations are introduced into each training mini-batch. This approach encourages inherent robustness to phase variations and indirectly mitigates performance degradation caused by the spatial offsets.

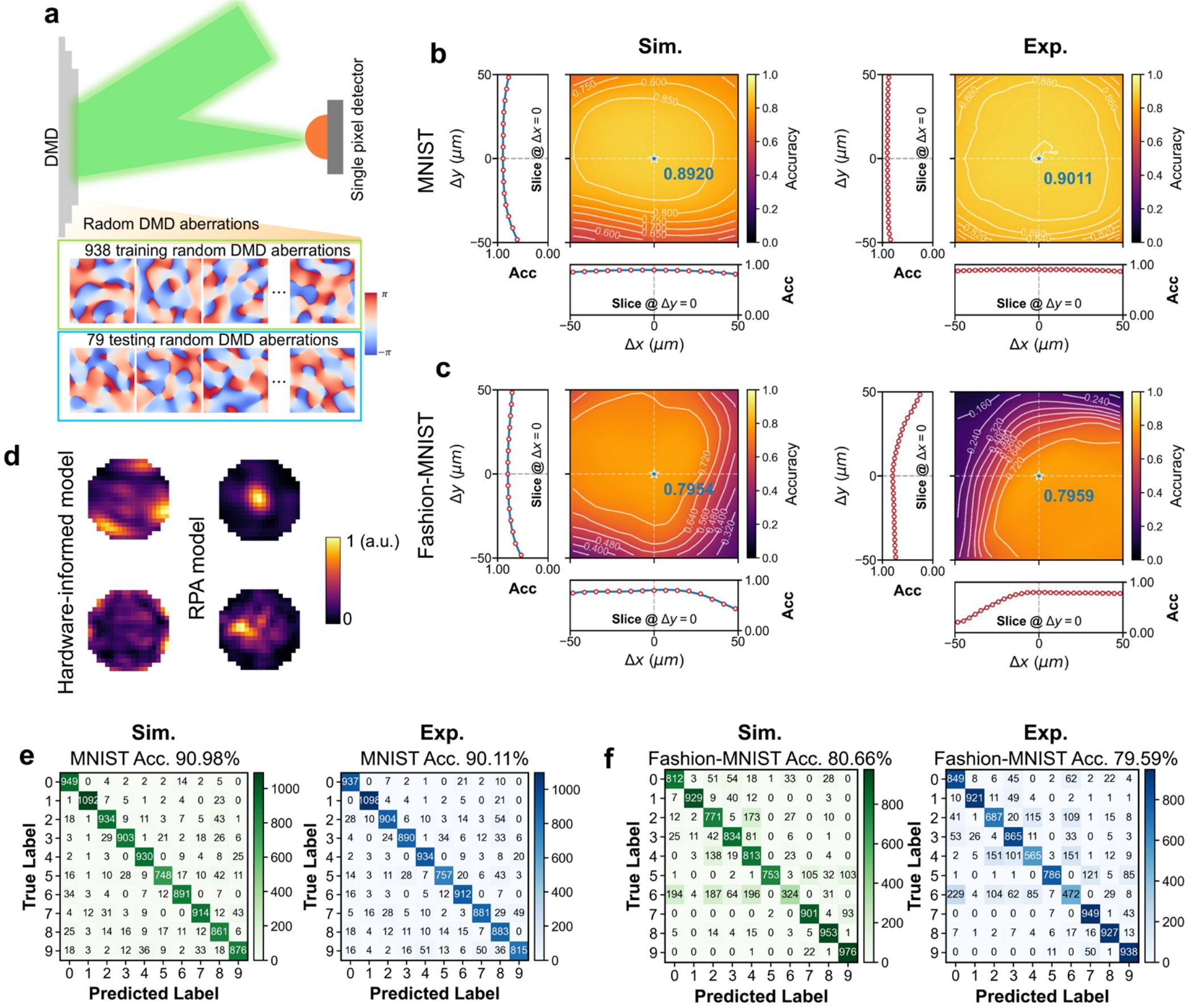

**Figure 4. Robustness enhancement and classification performance enabled by RPA**. **a**. Schematic of the RPA strategy. Instead of relying on a fixed DMD phase-aberration model, stochastic phase perturbations are injected at each training epoch, while validation is performed using phase profiles unseen during the training. **b**, **c**. Classification accuracy as a function of positional offsets $(\Delta x, \Delta y)$ for (**b**) MNIST and (**c**) Fashion-MNIST, for both numerical simulations and experimental measurements. **d.** Ensemble-averaged output-field intensity distributions for correctly classified samples under fixed-phase and RPA training. RPA produces a pronounced shift from peripheral divergence to strong energy aggregation near the center. **e, f.** Experimental confusion matrices for (e)MNIST and (f)Fashion-MNIST test sets, with peak accuracies up to around 90% and 80%, respectively.

Simulation results confirm that under randomized phase perturbations during testing, the model maintains high performance, reaching peak accuracies of ~90% and ~80% on MNIST and Fashion-MNIST, respectively. Figures 4b and 4c show the accuracy versus positional offset for the RPA-trained model on both datasets. In contrast to the sharp drops observed in Figs. 3d and 3e, the RPA-trained model exhibits markedly improved robustness, maintaining stable accuracy over a wide offset range. Even when the offset exceeds 26% of the detection region, accuracy remains over 88% on MNIST. Notably, although only phase perturbations are introduced during the training process, the model naturally develops implicit tolerance with respect to spatial offsets, as well. This can be attributed to the angular spectrum components with varying phase gradients introduced by RPA, which act as equivalent spatial translations during training and enable adaptive compensation for positional misalignment.

A statistical comparison of average intensity distributions (see Fig. 4d) between hardware-informed and RPA-trained models reveals a distinct shift toward central energy aggregation. While hardware-informed models show dispersed profiles with peripheral leakage, RPA-trained models concentrate energy at the detection core, enhancing classification reliability and broadening tolerance to both phase and positional errors. Figures 4e and 4f present confusion matrices for the RPA-trained model, with experimental accuracies being around 90% and 80% on MNIST and Fashion-MNIST validation sets, respectively. The simulated confusion matrix was evaluated under an aberration-free forward model to indicate the model's upper-bound performance. These results unambiguously demonstrate that RPA substantially enhances system robustness with quite minimal accuracy loss.

**Single-pixel Photodetector-Based Validation**

We further evaluated the system using single-pixel readout on a high-speed DMD platform operating at a 50-kHz refresh rate. Figures 5a and 5b show temporal signals captured by the single-pixel PD for representative MNIST and Fashion-MNIST samples. For each sample, the DMD sequentially displays a 10-frame composite pattern sequence. Two all-dark null frames are inserted between consecutive samples to provide clear temporal boundaries for visualization. Within each 10-frame sequence, the maximum response consistently occurs in the time window corresponding to the class-matched frame. Classification is thus performed by identifying the frame yielding the maximum stable response. Figure 5c shows a magnified view of the 10-frame temporal waveform for a Fashion-MNIST "pullover" sample, together with a zoom-in of a representative single-frame waveform. For each frame, the curve exhibits a characteristic "rise–plateau–fall" temporal profile, with the stable plateau window (shaded area) used for computing category scores. A zoom-in view of a single frame reveals a 20 μs period (50 kHz), comprising ~9 μs rising edge, ~7 μs stable plateau, and ~4 μs falling edge. For individual sample, we compute a category score for each of the ten frames by averaging the voltage within the 7 μs plateau window and infer the label using a maximum-response (argmax) rule. With a 50 kHz DMD refresh rate and ten frames per inference, the theoretical readout rate is 5 kHz. At this readout rate, the system achieves ~90% accuracy on the MNIST test set and ~80% on the Fashion-MNIST test set. The corresponding confusion matrices are shown in Fig. 5d. Given the small detection region (~137 μm) and high sensitivity to spatial offsets, precise micrometer level alignment of the pinhole in front of the PD is essential. This was achieved using an imaging system combined with holographic projection (Supplementary Note 4). We further assessed robustness through repeated recalibration-testing trials. Compared with the hardware-informed model trained solely on the measured phase aberration, the RPA model exhibited substantially higher robustness across trials (Supplementary Note 6).

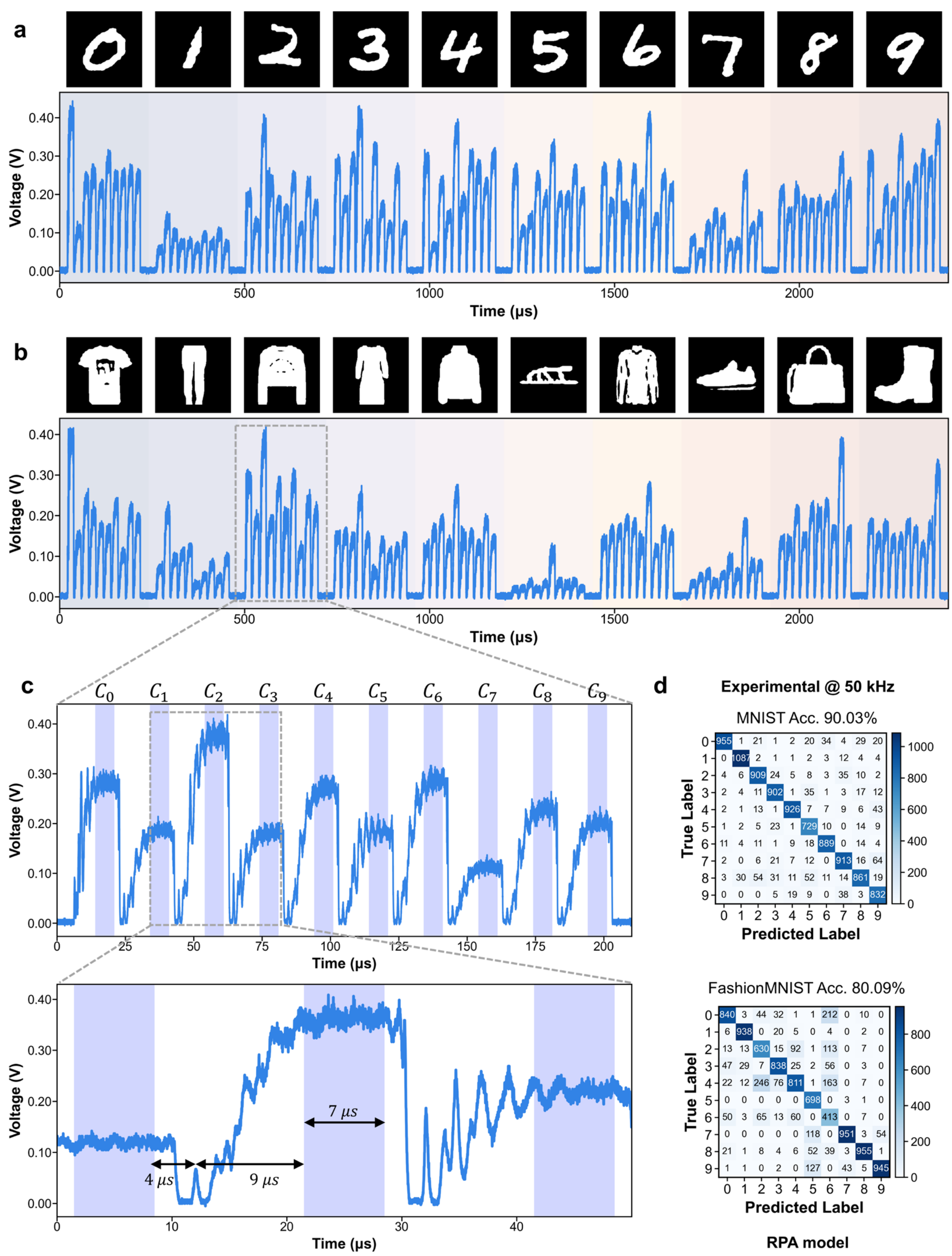

**Figure 5. High-speed single-pixel readout at 5 kHz.** A separate DLP7000 platform with a high-speed controller (50 kHz) was used for the measurements shown in this figure; see Methods for details. **a,b,** Representative MNIST and Fashion-MNIST samples, showing the input images and the corresponding PD temporal signals. For all samples, the maximum response consistently appears in the class-matched frame. **c,** Zoomed-in view of a Fashion-MNIST "pullover" waveform segment (top) and a representative single-frame trace (bottom). Each 20 µs frame (50 kHz) comprises an ~9 µs rising edge, an ~7 µs stable plateau, and an ~4 µs falling edge. **d,** Confusion matrices at a 5 kHz readout rate (50 kHz / 10 frames), showing ~90% accuracy on MNIST and ~80% on Fashion-MNIST.

## Discussions

This work introduces and experimentally validates a Gated-SP-DONN. In terms of inference logic, our system aligns naturally with the forward-forward algorithm (FFA). While FFA evaluates activation strength of input-label pairs, our architecture digitally combines the input image with pre-designed category-index masks in the time domain and assigns the category whose frame yields the highest detected intensity. Physically, each mask acts as a "virtual optical gate" at the detection region. When input features match a given category mask, the optical field concentrates energy into the detection region ("open" state); however, mismatched combinations scatter energy away, resulting in a weak response ("closed" state).

Compared with array detectors (e.g., CMOS, CCD), single-pixel detectors offer orders-of-magnitude higher bandwidth (up to GHz) and broader spectral coverage from ultraviolet to far-infrared. This circumvents the high cost and fabrication complexity of array detectors in non-visible bands, providing a cost-effective sensing solution for optical computing beyond the visible range. System throughput is governed by modulator refresh rate and detector bandwidth. While our prototype operates at 50 kHz using a DMD, the architecture is not intrinsically DMD-limited. Replacing it with ultrafast electrooptic modulators could push operational speeds into the MHz or even GHz regime. To demonstrate engineering feasibility, we have integrated a lightweight backend based on an STM32 minimal-system microcontroller to perform real-time classification of random handwritten digits (see Video 1), highlighting the potential for low-power, high-speed intelligent sensing.

We further note that the architecture is inherently scalable. Differential detection could further improve accuracy, while a hierarchical retrieval strategy, e.g., from coarse feature screening to fine-grained classification, could shorten the detection window. For instance, an initial stage using two masks for parity screening followed by five masks for category classification would probably reduce the required modulations from ten to seven, though this may involve a trade-off between speed and precision.

In summary, we have demonstrated a class-gated single-pixel diffractive neural network that compresses traditional multi-layer spatial architecture into a minimalist, time-multiplexed single-layer framework. By incorporating random phase augmentation during training, the system achieves robust performance under real optical imperfections. Combining ultra-compact architecture with high bandwidth single-pixel detection, this approach offers exceptional potential for real time sensing tasks such as high-speed target selection and capture of transient phenomena, paving a practical pathway for transitioning optical computing from laboratory demonstration to real-world deployment.

## Methods

### Experiment setup

The optical system is constructed using commercially available components (Fig. 2c). A 532 nm solid-state laser (Changchun New Industries Optoelectronics, MSL-III-532-50 mW) is coupled into a single-mode fiber for beam delivery. The output beam is collimated by a plano-convex lens (Hengyang, GLA11-050-150-A, focal length 150 mm), and collimation was verified with a shearing interferometer (LBTEK, TSI254), before being directed at a 24° incidence angle onto a DMD (Texas Instruments, DLP7000, 1024×768, 13.68 μm micromirror pitch) via a steering mirror (JCOPTIX, OME2-R2). All measurements used a DMD driven by an XD-ED01N controller (Xintong Technology), except the high-speed experiments shown in Fig. 5, which were acquired on a separate DLP7000 platform using a 50 kHz controller (FLDISCOVERY, F4320-DDR-0.7-XGA); a refresh rate of up to 50 kHz can be achieved when the number of lines in windowed playback is less than 337. Although both platforms employ the same DLP7000 chip family, differences in controller timing, effective device aberrations, and alignment conditions can produce minor variations in measured accuracy.

After modulation, the reflected light is relayed through a 4f system composed of two achromatic lenses: L1 (Daheng Optics, GCL-010617A, $f$ = 200 mm) and L2 (Daheng Optics, GCL-010604A, $f$ = 100 mm). An aperture placed at the Fourier plane suppresses higher order diffraction. A non-polarizing beam splitter divides the output into two paths. One path is directed to an industrial camera (HIKROBOT, MV-CH120-10UM, pixel size 3.45 μm, 12-bit depth; ROI-adjustable frame rate; $800 \times 800$ resolution yielding ~100 Hz) for recording the spatial light field distribution. The second path is focused onto a single-pixel PD (Thorlabs APD120, ~50 MHz bandwidth) for temporal intensity measurements. A 140 μm pinhole is placed directly in front of the detector to block out-of-region stray light. Camera alignment and region-of-interest (ROI) positioning are calibrated by loading a reference holographic mask onto the DMD. The PD output is recorded using an oscilloscope (Keysight, DSOX3024T, maximum sampling rate 200 MHz) set to a 20 MHz sampling rate to match the detector bandwidth. Synchronization between DMD frame switching and oscilloscope triggering is achieved via the DMD's output pulse signal.

**Optical Computational Model**

Our system implements a single reconfigurable diffractive layer using a DMD. For each input image $I(x_i, y_i)$, the system sequentially combines it with all class-specific masks $m_n(x_i, y_i)$ for $n \in \{0,1,\dots,9\}$. Each combination produces a composite pattern

$$C_n(x_i, y_i) = \big(I(x_i, y_i) \odot m_n(x_i, y_i)\big) \exp\,\big(j\,\phi(x_i, y_i)\big), \tag{1}$$

where $\odot$ denotes the Hadamard (element-wise) product. The phase term $\phi(x_i, y_i)$ specifies the model variant: $\phi(x_i, y_i) \equiv 0$ for the idealized model; $\phi(x_i, y_i) = \phi_{\mathrm{DMD}}^{\mathrm{measured}}(x_i, y_i)$ for the hardware-informed model; and $\phi(x_i, y_i) = \phi_{\mathrm{rand}}^{(k)}(x_i, y_i)$ for the RPA model, with $k$ indexing different random phase samples.

Because the DMD operates as a binary amplitude modulator, both input images and modulation masks are constrained to values in $\{0,1\}$. Images from the MNIST and Fashion-MNIST datasets are resized to the network resolution (256×256) and hard-quantized into binary patterns. However, to allow gradient-based optimization, we introduce a latent variable $\Theta_n(x_i, y_i) \in (-\infty, \infty)$and map it to a quasi-binary mask through a differentiable soft-quantization function:

$$m_n(x_i, y_i) = S(\Theta_n(x_i, y_i), \tau) = \frac{1}{1 + \exp\left(-\tau(\Theta_n(x_i, y_i) - 0.5\right)}, \tag{2}$$

where the temperature parameter $\tau$controls the steepness of the sigmoid and determines how closely the mask approximates a binary pattern. During training, $\tau$ is set to 1 during the first 10 epochs and then gradually increased to 110 over the remaining 10 epochs to obtain stable quasi-binary masks.

The composite pattern $C_n(x_i, y_i)$ is propagated over a distance $z$ to the output plane using the angular spectrum method:

$$U_n^{\text{out}}(x_i, y_i) = \mathcal{F}^{-1}\left[\mathcal{F}\left(C_n(x_i, y_i)\right) H\left(f_x, f_y; \lambda, z\right)\right], \tag{3}$$

where $\mathcal{F}$ and $\mathcal{F}^{-1}$denote the 2D Fourier transform and its inverse, and the transfer function is

$$H\left(f_x, f_y; \lambda, z\right) = \begin{cases} \exp\left[i2\pi z\sqrt{\frac{1}{\lambda^2} - f_x^2 - f_y^2}\right], & f_x^2 + f_y^2 < \frac{1}{\lambda^2}, \\ 0, & f_x^2 + f_y^2 \geq \frac{1}{\lambda^2}. \end{cases} \tag{4}$$

Here $f_x$ , $f_y$ are the spatial frequency coordinates corresponding to the $x$ and $y$ directions, and $\lambda$ is the wavelength. On the output plane, a circular detection region with radius $r = 10$ pixels is defined using the binary mask

$$\text{Mask}_r(x_i, y_i) = \begin{cases} 1, & x_i^2 + y_i^2 \leq r^2, \\ 0, & \text{otherwise.} \end{cases} \tag{5}$$

For a given data class $c$, the class score is computed by integrating the output intensity over this region:

$$s_c = \sum_{x_i, y_i} \text{Mask}_r\,(x_i, y_i)\, I_c^{\text{out}}(x_i, y_i). \tag{6}$$

The class with the highest score is taken as the final inference result.

In the single-pixel readout experiment, the PD signal $I_{\text{PD}}(t)$ is divided into 10 temporal segments according to the DMD frame period. For class $c$, the response is obtained by averaging the samples within a stable interval $\left[t_c^{\text{start}},\, t_c^{\text{end}}\right]$:

$$s_c^{\text{PD}} = \frac{1}{K}\sum_{k=1}^{K} I_{\text{PD}}\,(t_k), t_k \in \left[t_c^{\text{start}}, t_c^{\text{end}}\right]. \tag{7}$$

The same maximum-response criterion as in spatial readout is used to determine the predicted label. The network is trained to maximize classification accuracy using a mean-squared error (MSE) loss applied to the SoftMax-normalized class scores:

$$L = \mathrm{MSE}\,(\sigma(\mathrm{s}), \mathbf{y}_{\mathrm{true}}), \tag{8}$$

where $\sigma(\cdot)$ denotes the SoftMax operator, $\mathbf{s} = [s_0, s_1, \dots, s_9]^\top$ is the class-score vector, and $\mathbf{y}_{\mathrm{true}}$is the one-hot ground-truth label. During training, mask parameters are represented by the latent variables $\Theta_n(x, y)$ and updated via backpropagation. After training, a Heaviside step function is applied to binarize the learned masks, converting the latent variables into physical binary patterns in $\{0,1\}$.

The network is configured with a spatial resolution of $256 \times 256$. Images from the MNIST and Fashion-MNIST datasets are resized from $28 \times 28$ to $256 \times 256$ and binarized before training. Mini-batches of 64 randomly sampled images are used for optimization. To enhance robustness against real-world optical perturbations, random phase disturbances are applied at the input plane during RPA training. Diffractive propagation is computed with the angular spectrum method. Model parameters are updated with the Adam optimizer at a learning rate of 0.1. Training is performed for 20 epochs: the temperature parameter $\tau$ is fixed at 1 during the first 10 epochs and then gradually increased to 110 over the remaining 10 epochs to obtain stable quasi-binary masks. All simulations are implemented in Python 3.9.1 and PyTorch 2.1.1 on a workstation equipped with an NVIDIA GeForce RTX 5090 GPU (32 GB VRAM) and an Intel® Core™ Ultra7-265K CPU.

**Supporting Information**

Supporting Information is available from the author.

**Acknowledgements**

This work was supported by National Natural Science Foundation of China (No. 62375064), Shenzhen Science and Technology Program (No. JCYJ20250604145307010, ZDCY20250901100959001, KJZD20230923114803007), and the National Key Research and Development Program of China (No. 2022YFB3603204).

**Conflict of Interest**

The authors declare no conflict of interest.

**Data Availability Statement**

The data that support the findings of this study are available from the corresponding author upon reasonable request.